\documentclass[10pt,final,twocolumn,journal]{IEEEtran}
\usepackage{filecontents,lipsum}
\usepackage[noadjust]{cite}
\usepackage[percent]{overpic}
\usepackage{stfloats}
\usepackage{ctable}
\usepackage[mathlines]{lineno}

\usepackage{cite}
\usepackage{graphicx}
\usepackage{caption}
\usepackage{subcaption}
\usepackage{authblk}
\usepackage{xcolor}
\usepackage{amsmath}
\usepackage{array}
\usepackage[normalem]{ulem}

\newcommand*{\Scale}[2][4]{\scalebox{#1}{$#2$}}%
\begin{document}
{\copyright\ 20xx IEEE. Personal use of this material is permitted. Permission
from IEEE must be obtained for all other uses, in any current or future
media, including reprinting/republishing this material for advertising or
promotional purposes, creating new collective works, for resale or
redistribution to servers or lists, or reuse of any copyrighted
component of this work in other works.}
\newpage
\title{Scaling Theory of Electrically Doped 2D Transistors}
\author{Hesameddin Ilatikhameneh, Gerhard Klimeck, Joerg Appenzeller, and Rajib Rahman
\thanks{This work was supported in part by the Center for Low Energy Systems Technology (LEAST), one of six centers of STARnet, a Semiconductor Research Corporation program sponsored by MARCO and DARPA.}
\thanks{The authors are with the Department
of Electrical and Computer Engineering, Purdue University, West Lafayette, IN, 47907 USA e-mail: hesam.ilati2@gmail.com.}
}
%



\maketitle
\setlength{\textfloatsep}{7pt plus 1.0pt minus 2.0pt}
\begin{abstract}

In this letter, it is shown that the existing scaling theories for chemically doped transistors cannot be applied to the novel class of electrically doped 2D transistors and the concept of equivalent oxide thickness (EOT) is not applicable anymore. Hence, a novel scaling theory is developed based on analytic solutions of the 2D Poisson equation. Full band atomistic quantum transport simulations verify the theory and show that the critical design parameters are the physical oxide thickness and distance between the gates. Accordingly, the most optimized electrically doped devices are those with the smallest spacing between the gates and the thinnest oxide, and not the smallest EOT.
\end{abstract}

\section{Introduction}
Ultra-thin channels are crucial in aggressively scaled field-effect transistors (FETs) and in tunnel FETs (TFETs) for better gate control and higher performance \cite{ITRS1, ITRS3, Appenzeller1}. The emergence of a novel class of 2D semiconducting materials such as phosphorene, silicine, and transition metal dichalcogenides (TMDs) has enabled the possibility of atomically thin channels in transistors \cite{Fiori, MoS2_Nature}. For example, monolayer MoS$_2$ transistors are shown to outperform Si FETs by a factor of 3 in terms of on-current and on/off ratio in 5nm channels \cite{Lake}. Moreover, 2D homojunction TMD TFETs can potentially provide high on-currents compared to other TFETs \cite{Hesam1}. However, to operate as TFETs, these materials need to be highly doped to realize tunnel junctions with high internal electric fields \cite{Hesam1}. 
Electrical doping methods wherein pn junction like band profiles are realized with multiple gates are both promising and the currently dominant doping method in 2D material transistors \cite{elec1, elec2}. Electrical doping also eradicates the deleterious effects of chemical doping in TFETs in which dopant states in the band gap can increase the off-currents \cite{sapan}.
Despite the large interest in electrically doped 2D devices and their promise for future electronics, there is no design guideline for these devices.  The well-known scaling theory \cite{Scaling1, Scaling2} can be used in \emph{conventional} transistors to quantify the effect of gate control on the electric field at the junctions \cite{Scaling_TFET}. However, atomistic simulations show that this theory fails to explain the behavior of \emph{electrically doped} 2D material transistors correctly. According to the conventional scaling theory, devices with similar equivalent oxide thicknesses (EOT) have similar scaling length, potential profile, and performance. However, simulations of WSe$_2$ TFET reveal that the devices with the same EOT but different oxide thicknesses can result in more than 3 orders of magnitude change in on-current.

In this work, a novel scaling theory is developed to help with the design optimization of \emph{electrically doped} devices for the first time. The scaling length ($\lambda$) of conventional FETs ($\lambda_{old}$) and electrically doped devices ($\lambda_{new}$) are compared in (1) and (2). According to the new scaling theory, high-k dielectrics are not necessarily favorable in electrically doped devices, but ultra-thin oxides with large band gaps to suppress the gate leakage are. Given the interest in 2D transistors, the theory presented here shows that a change in the design principles is needed when TFETs are electrically doped. {Notice that this reduced impact of the EOT is essentially due to the presence of \emph{fringing fields} and it is not limited to 2D devices only. In the devices in which fringing fields are strong the impact of the oxide dielectric constant reduces \cite{EOT}.
\begin{equation}
\label{eq:scaling2}
\lambda_{old} \approx \sqrt{ \frac{1}{2} \frac{t_{ox}}{\epsilon_{ox}} \epsilon_{ch}t_{ch} (1+\frac{\epsilon_{ox}}{t_{ox}} \frac{t_{ch}}{4\epsilon_{ch}})} = f(EOT) \\
\end{equation}
}
\begin{equation}
\label{eq:scaling7}
\lambda_{new} \approx \frac{2 t_{ox} + S/4}{\pi} = f(t_{ox})\\
\end{equation}

\setlength{\belowdisplayskip}{4pt} \setlength{\belowdisplayshortskip}{4pt}
\setlength{\abovedisplayskip}{4pt} \setlength{\abovedisplayshortskip}{4pt}

\noindent Here, $S$ is the distance between the gates, $\epsilon_{ch}$, and $\epsilon_{ox}$ are the dielectric constants of channel and oxide respectively, whereas $t_{ch}$, and $t_{ox}$ are their thicknesses. 
%

First, full-band quantum transport simulations are performed on a monolayer 2D TMD WSe$_2$ homojunction TFET in Section III to show the critical design parameters involved in electrical doping. Based on the simulation results, the inadequacies of the original scaling theory are then highlighted, and a new analytical theory is developed in Section IV.  

%


\section{Simulation details}

The atomistic quantum transport simulations use a sp$^3$d$^5$ 2nd nearest neighbor tight-binding model and a self-consistent Poisson-NEGF (Non-equilibirum Green's Function) method. {Neumann boundary conditions are used for the Poisson equation at the source and drain boundaries and the source and drain contact voltages determine the Fermi levels of the contacts}. The details of the models and the methods can be found in \cite{Hesam1}. {The material properties of monolayer WSe$_2$ are listed in Table \ref{tab:tmd_prop}.} The simulated monolayer WSe$_2$ TFET assumes a structure shown in Fig. \ref{fig:Fig1}a. Each gate has a length of 13nm, with the gate spacing $S$ assumed to be zero unless otherwise stated. An electrically doped pn junction can be created in this structure by applying biases of opposite polarity in the two gates. A source-drain voltage $V_{DS}$ of 1V is used throughout, and the relative dielectric constant and physical oxide thickness are set to 20 and 1.66nm respectively unless mentioned otherwise. The total thickness of the device (the distance between the top and bottom gates shown as $T_{tot}$ in Fig. \ref{fig:Fig1}a) equals 4nm by default, including the body thickness of monolayer WSe$_2$ ($\approx$0.67nm). All of the transport simulations have been performed with the simulation tool NEMO5 \cite{nemo5_1}.

{\captionsetup[table]{belowskip=10pt,aboveskip=15pt}
\begin{table}[h!]\center
\caption{\label{tab:tmd_prop} {WSe$_2$ material properties: band gap ($E_g$), electron and hole effective masses ($m^*_e$ and $m^*_h$), and in-plane and out-of-plane relative dielectric constants ($\epsilon_r^{in}$ and $\epsilon_r^{out}$) \cite{eps_tmds}.}}
    \begin{tabular}{| l | l | l | l | l | l |}
    \hline
	Parameters & $E_g$ [eV] & $m^*_e$ [m$_0$] & $m^*_h$ [m$_0$] & $\epsilon_r^{in}$ & $\epsilon_r^{out}$   \\ \hline
	 WSe$_2$ & 1.56 & 0.36 & 0.5 & 4.5 & 2.9\\
	\hline
    \end{tabular}
\end{table}

\begin{figure}[h!]
        \centering
        \begin{subfigure}[b]{0.45\textwidth}
               \includegraphics[width=\textwidth]{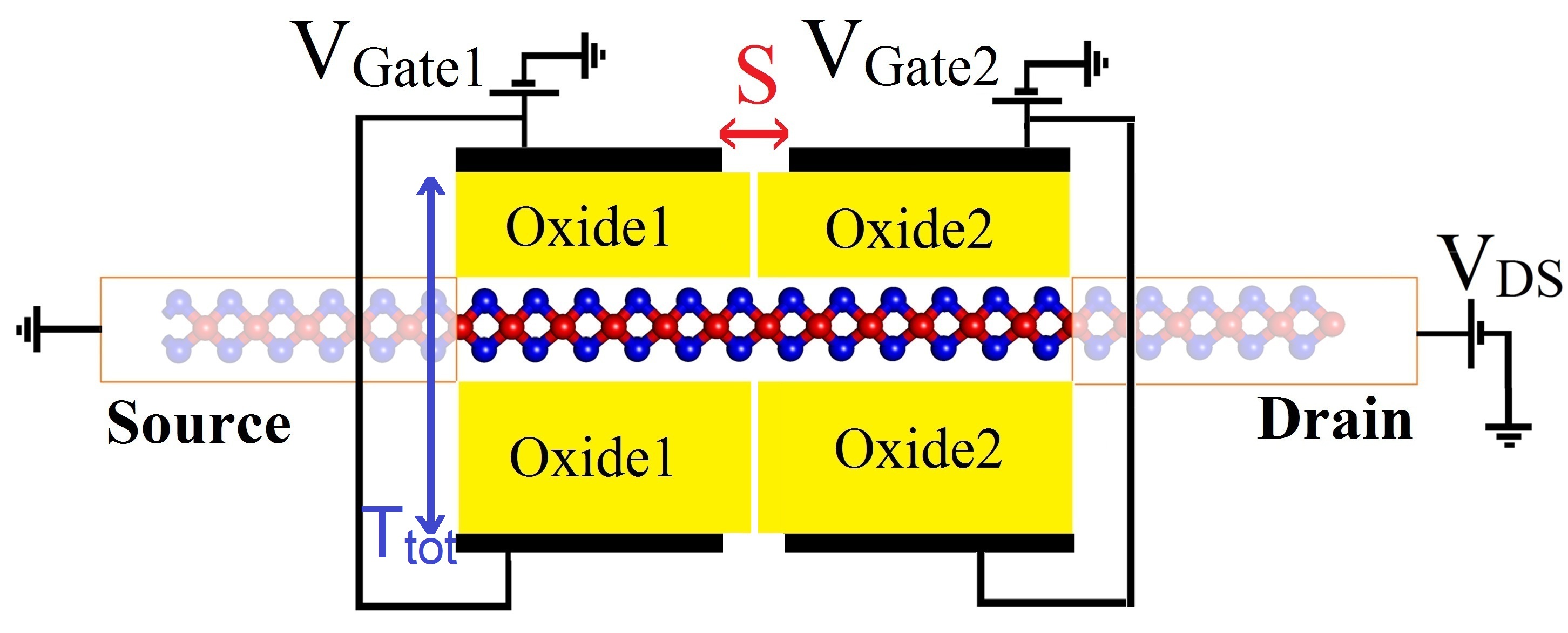}
               \vspace{-2.4\baselineskip}
                \caption{}
                \label{fig:struct1}
        \end{subfigure}%
        \quad
        \begin{subfigure}[b]{0.25\textwidth}
               \includegraphics[width=\textwidth]{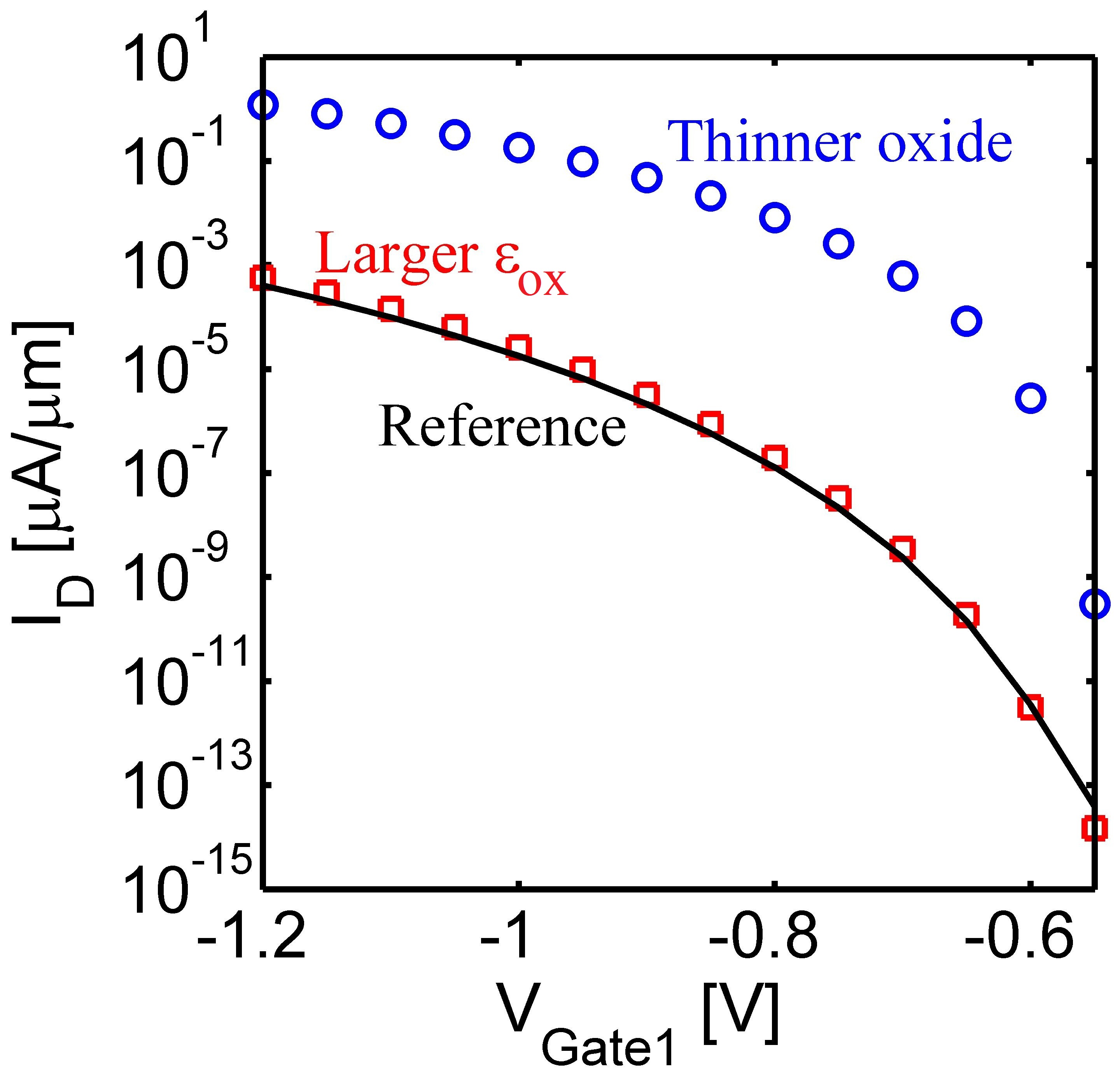}
               \vspace{-1.5\baselineskip}
                \caption{}
                \label{fig:Same_EoT}
        \end{subfigure}%
        \begin{subfigure}[b]{0.247\textwidth}
               \includegraphics[width=\textwidth]{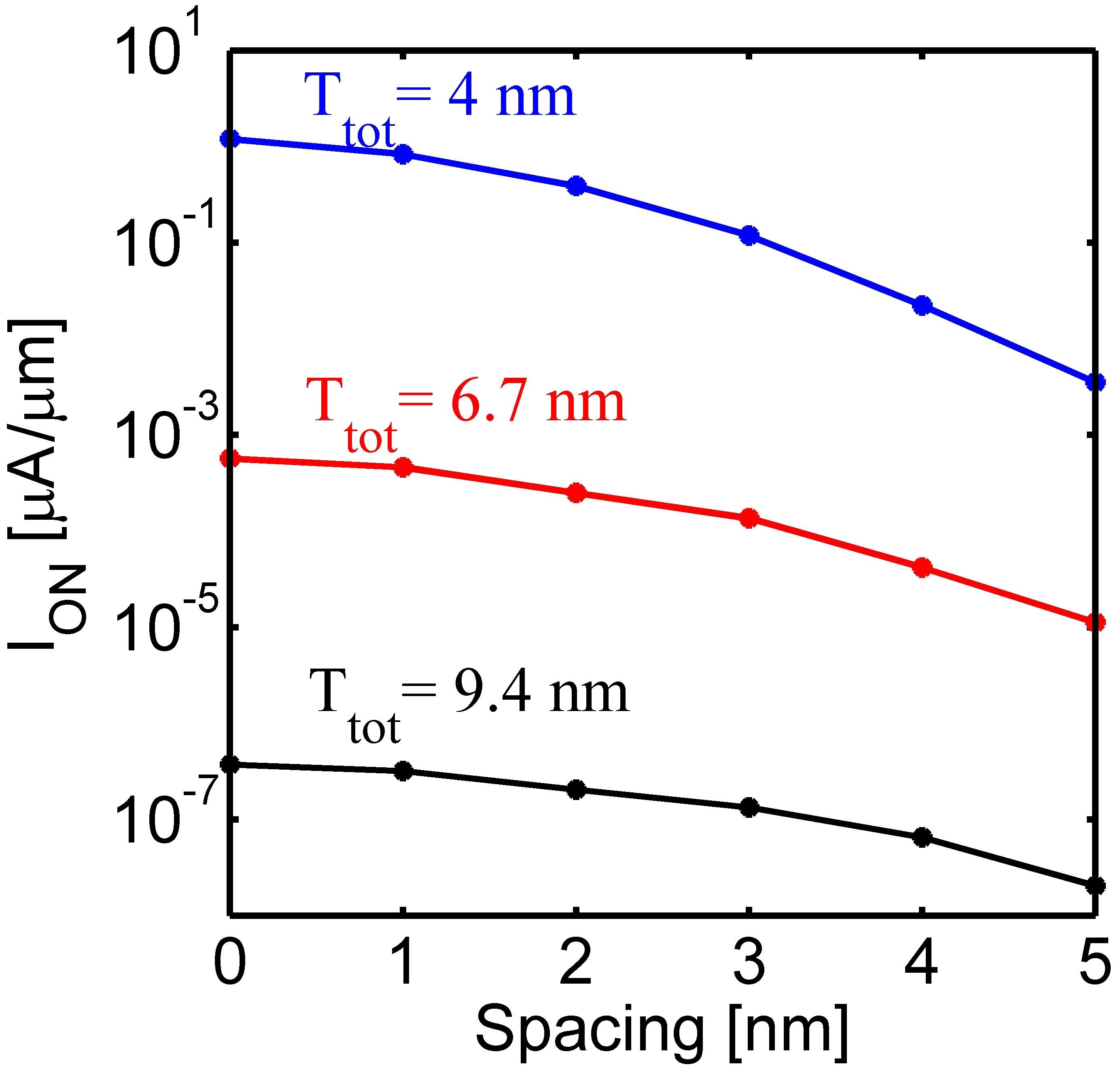}
               \vspace{-1.5\baselineskip}
                \caption{}
                \label{fig:Spacing}
        \end{subfigure}%
        \vspace{-0.7\baselineskip}
        
        \caption{a) Physical structure of an electrically doped mono-layer WSe$_2$ TFET with left and right oxide lengths of 13nm. {$\rm V_{Gate2}$ is fixed to 1V. b) Transfer characteristics of the TFET with $T_{tot}=6.7nm$ and $\epsilon_{ox}=11$ as reference (black line), with $T_{tot}=6.7nm$ and $\epsilon_{ox}=20$ (red squares), and with $T_{tot}=4nm$ and $\epsilon_{ox}=11$ (blue circles),} c) On-current of the TFET with gate spacing $S$.}\label{fig:Fig1}
\end{figure}

\section{Simulation results}





Fig. \ref{fig:Fig1}b shows the transfer characteristics of an electrically doped TFET. {Notice that the tunneling transmission depends exponentially on the electric field at the tunnel junction and a small variation in the electric field makes a large difference in the on-current. The I-V curve labeled as reference belongs to a $T_{tot}$ of 6.7nm and $\epsilon_{ox}$ of 11. There are two ways to decrease EOT: 1) increase $\epsilon_{ox}$ or 2) decrease the thickness of the oxide. } According to the previous scaling theory, these two methods should lead to similar results since they result in the same EOT. {However, the numerical results show that changing the dielectric constant does not affect the I-V significantly (increase in the on-current by a factor of 1.4), while decreasing the physical thickness of the oxide does (increase in the on-current by a factor of 3000).} This is a critical finding of the new scaling theory, i.e. that the main scaling parameter is the physical thickness of the oxide. This implies that to obtain better performance in an electrically doped 2D TFET, the physical thickness of the oxide should be reduced, not just EOT.

Fig. \ref{fig:Fig1}c shows the effect of gate spacing on the on-current of the WSe$_2$ TFET. Increasing the spacing reduces the on-current significantly, however, this effect is not as large as the effect of increasing the oxide thickness. Numerical simulations therefore identify the physical oxide thickness and the gate spacing as two critical parameters for electrical doping. This is essentially due to the presence of fringing fields from the gates. As long as these fringing fields are strong, the physical thickness of the oxide matters more than the EOT.
\section{Analytic Modeling}
\label{sec:analytic_sec}
To find an analytic solution of the device potential due to electrical doping, four assumptions have been made here: 1) there is no spacing ($S=0$) between the gates but the gates are electrically isolated, 2) the thickness of the 2D material is neglected compared to the {device thickness: $t_{ch} \ll T_{tot}$} (or the dielectric constant of the 2D material is close to that of the oxide), 3) the length of the device is much larger than its thickness {($L_{Ch} \gg T_{tot}/\pi$)}, 4) the mobile charge within the tunnel junction is small enough to be neglected in the Poisson equation such that the Laplace equation can be solved instead \cite{Scaling_TFET}. Later on, assumption 1) is relaxed by introducing an empirical parameter $\alpha$ in the analytic model that captures the effect of spacing on the scaling length $\lambda$. Assumptions 2) and 3) are justified from the physical structure of 2D material FETs and their sub-nanometer thicknesses \cite{MoS2_Nature}. Assumption 4) is valid if the magnitude of the mobile charge $n$ is much smaller than a critical charge $n_0$ {($n \ll n_{0}$)}, which depends on the magnitude of the second derivatives of the potential close to the tunnel junction, $n_0 \approx \frac{\epsilon |V_1-V_2|}{q\lambda^2}$. In a typical case, for $\lambda$ of 1nm, oxide relative dielectric constant $\epsilon_{ox}$ of 10, and a potential difference between the gates ($V_1-V_2$) of 1V, $n_0$ is about $5e20 cm^{-3}$. The high value of $n_0$ makes assumption 4 valid for most practical situations including graphene. It is to be noted that the numerical simulations do not make any of these assumptions. Hence, a comparison between the two will justify the analytic model. Considering the above assumptions, the 2D Laplace equation can be solved with the boundary conditions shown in Fig. \ref{fig:Fig2}a to obtain the potential profile inside the device.
\begin{equation}
\label{eq:new_scaling1}
\Scale[0.95]{
V(x,y)=\sum\limits_{k=0}^\infty A_k {sin}\left(\frac{(2k+1)\pi}{T_{tot}}y \right) {exp}\left(-\frac{(2k+1)\pi}{T_{tot}}  x\right) + C\\
}
\end{equation}

\noindent
Since the first term of the series ($k=0$) in (\ref{eq:new_scaling1}) has the largest magnitude and the smallest slope, it is the limiting factor for the magnitude of the electric field. Hence, considering this term only in (\ref{eq:new_scaling1}), the potential along the channel ($y = T_{tot}/2$) is given as,
\begin{equation}
\label{eq:new_scaling2}
\Scale[0.84]{
V(x,\frac{T_{tot}}{2})\approx 
\begin{cases}
(V_M - V_1) {exp}\left(+\frac{\pi}{T_{tot}} (x-x_M)\right) + V_1 & x<x_M \\
(V_M - V_2) {exp}\left(-\frac{\pi}{T_{tot}} (x-x_M)\right) + V_2 & x>x_M \\
\end{cases}}
\end{equation}
\noindent
where $V_M$ and $x_M$ are the potential and the position of the interface between the gates respectively. Using continuity of the displacement field at $(x,y)=(0,T_{tot}/2)$, $V_M\approx \frac{\epsilon_1 V_1+\epsilon_2 V_2}{\epsilon_1+\epsilon_2 }$, for the general asymmetric gating case shown in Fig. \ref{fig:Fig2}a with two different dielectric constants $\epsilon_1$ and $\epsilon_2$ for the p and n junctions. For the case of symmtric gating with $\epsilon_1=\epsilon_2$, $V_M = (V_1+V_2)/2$. Comparing the above solution with a conventional scaling theory solution $exp(-x/\lambda)$, the natural scaling length can be extracted as $\lambda \approx T_{tot}/\pi$, which shows again that the scaling length depends on the physical oxide thickness, and not on EOT. 




Fig. \ref{fig:Fig2}b depicts the potential profile along the channel obtained from NEGF simulations and the approximate solution using (\ref{eq:new_scaling2}). 
The analytical potential profile matches the numerical results very well for different physical oxide thicknesses which supports the validity of the new scaling length. Fig. \ref{fig:Fig2}c shows the conduction band profile of the pn junction with different asymmetric gating ratios $\epsilon_1/\epsilon_2 = 4, 1,$ and $1/4$ showing excellent agreement again between the analytic results and NEGF. Notice that reducing or increasing $\epsilon$ of the right gate oxide ($\epsilon_2$) compared to the left ($\epsilon_1$) does not change the electric field significantly. {This analytic model is also applied to other materials such as bilayer graphene and monolayer WTe$_2$ and verified by comparison with full-band atomistic simulations.}


\begin{figure}[h!]
        \centering
        \begin{subfigure}[b]{0.3\textwidth}
               \includegraphics[width=\textwidth]{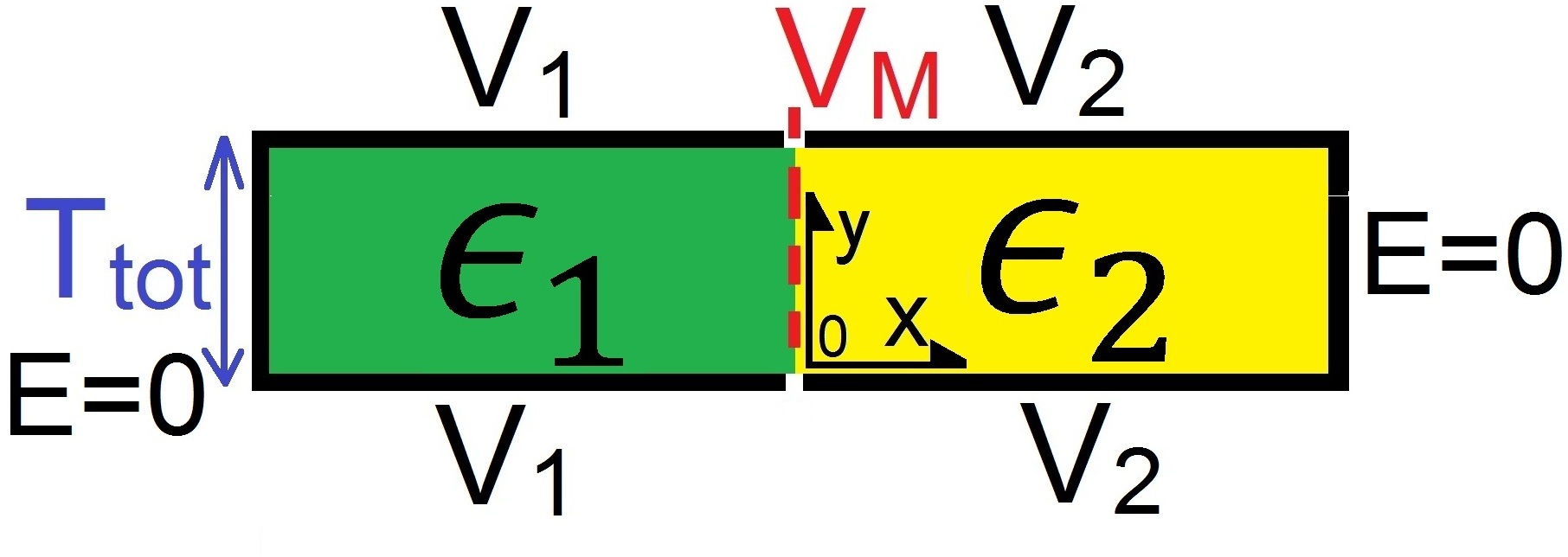}
               \vspace{-2.5\baselineskip}               
                \caption{}
                \label{fig:laplace}
        \end{subfigure}%
        \quad        
        \begin{subfigure}[b]{0.47\textwidth}
               \includegraphics[width=\textwidth]{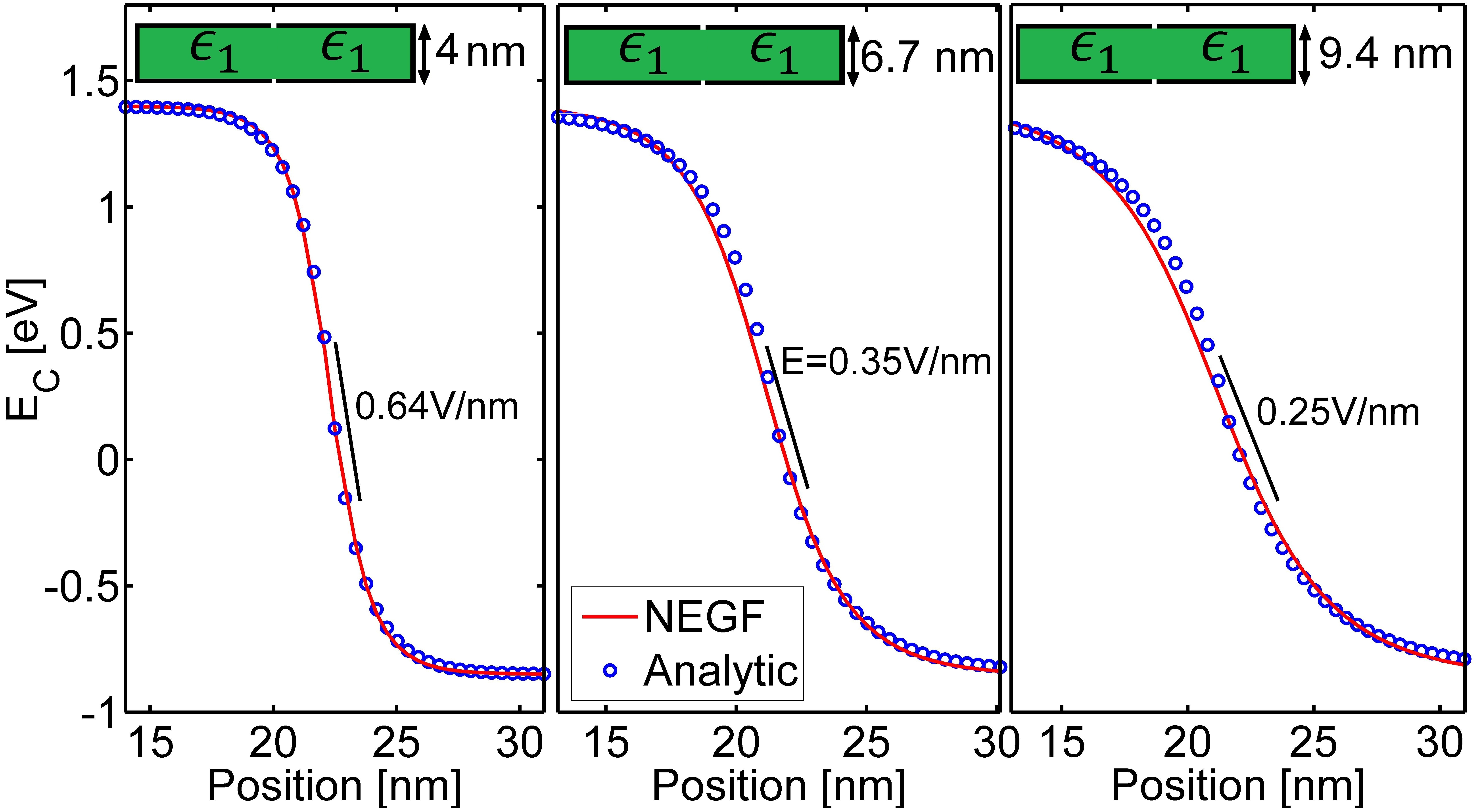}
               \vspace{-1.5\baselineskip}                              
                \caption{}
                \label{fig:st3}
        \end{subfigure}%
        \quad            
        \begin{subfigure}[b]{0.47\textwidth}
               \includegraphics[width=\textwidth]{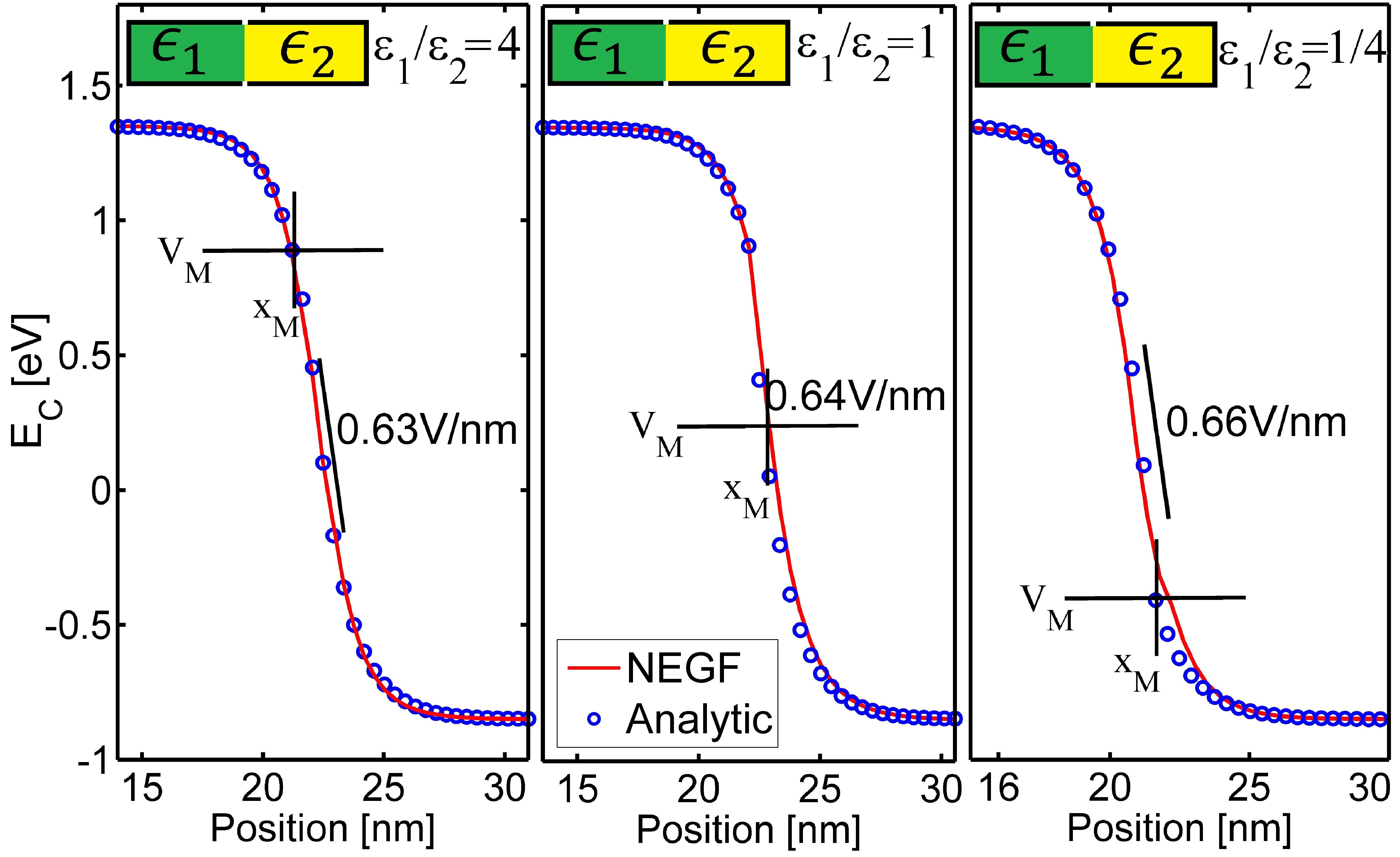}
               \vspace{-1.5\baselineskip}                              
                \caption{}
                \label{fig:st3}
        \end{subfigure}%
        \vspace{-0.5\baselineskip}       
        \caption{a) Domain and boundary conditions of the simplified problem for the purpose of finding an analytic solution, b) The potential profile along the channel obtained from atomistic simulation (red lines) and analytic expressions (blue circles) for different total physical thicknesses of the device and c) various $\epsilon_1 / \epsilon_2$ values. {$|V_1-V_2|$ equals $2.2V$ here.} }\label{fig:Fig2}	
\end{figure}

An empirical factor $\alpha$ can be used to capture the effect of spacing $S$ on the scaling length, as given by 
\begin{equation}
\label{eq:new_scaling5}
\lambda \approx \frac{T_{tot} + \alpha S}{\pi} \\
\end{equation}
From atomistic simulation results, $\alpha$ is found to be about 1/4, which implies that $T_{tot}$ has more impact on the performance of the device if compared to the spacing $S$ (since $\alpha<1$). {This value of $\alpha$ ($\alpha$ = 0.25) is obtained and validated for spacing distances in the range of 0 to 5nm.}


\section{Conclusion}
In this work, a new scaling theory is developed for electrically doped 2D transistors. The predictions of the new theory differ significantly from those of the old scaling theory, and are justified by full band atomistic NEGF simulations. 
The major players in the performance of electrically doped 2D transistors are found to be the physical thickness of the oxides and the spacing distance between the gates. Among the two, the spacing has less impact by a factor of about 1/4. As a result, it is critical for electrically doped 2D TFETs to reduce the physical thickness of oxide and fabricate the gates as close as possible for high performance applications.

\end{document}